\documentclass[reprint,superscriptaddress,aps,prb,twocolumn,floatfix]{revtex4-2}
\usepackage[utf8]{inputenc}
\usepackage{textcomp} 
\usepackage{setspace}
\usepackage{amsmath}
\usepackage{breqn}
\usepackage{graphicx}
\usepackage{verbatim}
\renewcommand{\figurename}{\textbf{Fig.}}
\usepackage{amsfonts}
\usepackage{amssymb}
\usepackage{xcolor}
\usepackage{soul}
\usepackage{tabularx}
\setcitestyle{super}
\usepackage{float}
\DeclareGraphicsExtensions{.pdf,.eps,.png,.jpg,.mps}

\begin{document}

\title{Hertz-linewidth semiconductor lasers using CMOS-ready ultra-high-$Q$ microresonators}

\author{
Warren Jin$^{1,\ast}$, Qi-Fan Yang$^{2,\ast}$, Lin Chang$^{1,\ast}$, Boqiang Shen$^{2,\ast}$, Heming Wang$^{2,\ast}$,\\Mark A. Leal$^{1}$, Lue Wu$^{2}$, Avi Feshali$^{3}$, Mario Paniccia$^{3}$, Kerry J. Vahala$^{2,\dagger}$, and John E. Bowers$^{1,\dagger}$\\
$^1$ECE Department, University of California Santa Barbara, Santa Barbara, CA 93106, USA\\
$^2$T. J. Watson Laboratory of Applied Physics, California Institute of Technology, Pasadena, CA 91125, USA\\
$^3$Anello Photonics, Santa Clara, CA\\
$^\ast$These authors contributed equally to this work.\\
$^\dagger$Corresponding authors: vahala@caltech.edu, bowers@ece.ucsb.edu\\
}

\begin{abstract}
\noindent{\bf Driven by narrow-linewidth bench-top lasers, coherent optical systems spanning optical communications, metrology and sensing provide unrivalled performance. To transfer these capabilities from the laboratory to the real world, a key missing ingredient is a mass-produced integrated laser with superior coherence. Here, we bridge conventional semiconductor lasers and coherent optical systems using CMOS-foundry-fabricated microresonators with record high $Q$ factor over 260 million and finesse over 42,000. Five orders-of-magnitude noise reduction in the pump laser is demonstrated, and for the first time, fundamental noise below 1~Hz$^2$~Hz$^{-1}$ is achieved in an electrically-pumped integrated laser. Moreover, the same configuration is shown to relieve dispersion requirements for microcomb generation that have handicapped certain nonlinear platforms. The simultaneous realization of record-high $Q$ factor, highly coherent lasers and frequency combs using foundry-based technologies paves the way for volume manufacturing of a wide range of coherent optical systems.}
\end{abstract}
\maketitle

The benefits of high coherence lasers extend to many applications. Hertz-level linewidth is required to interrogate and manipulate atomic transitions with long coherence times, which form the basis of optical atomic clocks \cite{ludlow2015optical,newman2019architecture}. Furthermore, linewidth directly impacts performance in optical sensing and signal generation applications, such as laser gyroscopes \cite{gundavarapu2019sub,lai2020earth}, light detection and ranging (LIDAR) systems \cite{trocha2018ultrafast,suh2018soliton}, spectroscopy \cite{suh2016microresonator}, optical frequency synthesis \cite{spencer2018optical}, microwave photonics \cite{Li2013,liang2015high,hao2018toward,marpaung2019integrated,liu2020photonic}, and coherent optical communications\cite{kikuchi2015fundamentals, olsson2018probabilistically}. In considering the future transfer of such high coherence technologies to a mass manufacturable form, semiconductor laser sources represent the most compelling choice. They are directly electrically pumped, wafer-scale manufacturable and capable of complex integration with other photonic devices. Indeed, their considerable advantages have made them into a kind of `photonic engine' for nearly all modern day optical source technology, including commercial benchtop laser sources. Nonetheless, mass manufacturable semiconductor lasers, such as used in communications systems, have linewidths ranging from 100~kHz to a few MHz \cite{kikuchi2015fundamentals}, which is many orders of magnitude too large for the above mentioned applications.

A powerful method to narrow the linewidth of a laser is to apply optical feedback through an external reflector, for which the degree of noise suppression scales with the square of the Q factor of the reflector \cite{hjelme1991semiconductor,liang2015ultralow,kondratiev2017self}. Ultra-high-$Q$ microresonators are excellent candidates to achieve substantial linewidth narrowing and have been demonstrated across a wide range of materials as discrete \cite{liang2015ultralow,lee2012chemically} or integrated components \cite{adar1994less,biberman2012ultralow,spencer2014integrated,ji2017ultra,zhang2017monolithic,yang2018bridging,gundavarapu2019sub,puckett2019silicon,liu2020photonic}. While sub-Hertz fundamental linewidth has been realized in semiconductor lasers that are self-injection-locked to discrete crystalline microresonators \cite{liang2015ultralow}, retaining ultra-high $Q$ factor when moving to higher levels of integration is both of  paramount importance and challenging. As  a measure of the level of difficulty,  current demonstrations of narrow-linewidth integrated lasers, despite many years of effort, feature fundamental linewidths of 40~Hz to 1~kHz, as limited by their $Q$ factors \cite{fan2019ultra,raja2019electrically,shen2020integrated,tran2019tutorial}.

In this work, we present critical advances in silicon nitride waveguides, fabricated in a high-volume complementary metal-oxide-semiconductor (CMOS) foundry. We achieve a $Q$ factor over 260 million -- a record among all integrated resonators. By self-injection locking a conventional semiconductor distributed-feedback (DFB) laser to these ultra-high-$Q$ microresonators, we reduce noise by five orders of magnitude, yielding frequency noise below 1~Hz$^2$~Hz$^{-1}$, which is a previously unattainable level for integrated lasers. Within the same configuration, a new regime of Kerr comb operation in microresonators is supported. Specifically, the comb both operates turnkey \cite{shen2020integrated} and attains coherent comb operation under conditions of normal dispersion without any special dispersion engineering.  The comb's line spacing is suitable for dense (DWDM) communications systems. Moreover, each comb line benefits from the exceptional frequency noise performance of the disciplined pump, representing a significant advance for DWDM source technology. The microwave phase noise performance of the comb is also comparable to that of existing commercial microwave oscillators. Overall, experiment and theory reveal an ultra-low-noise regime in integrated photonics.

\begin{figure*}
\centering
\includegraphics[width=\linewidth]{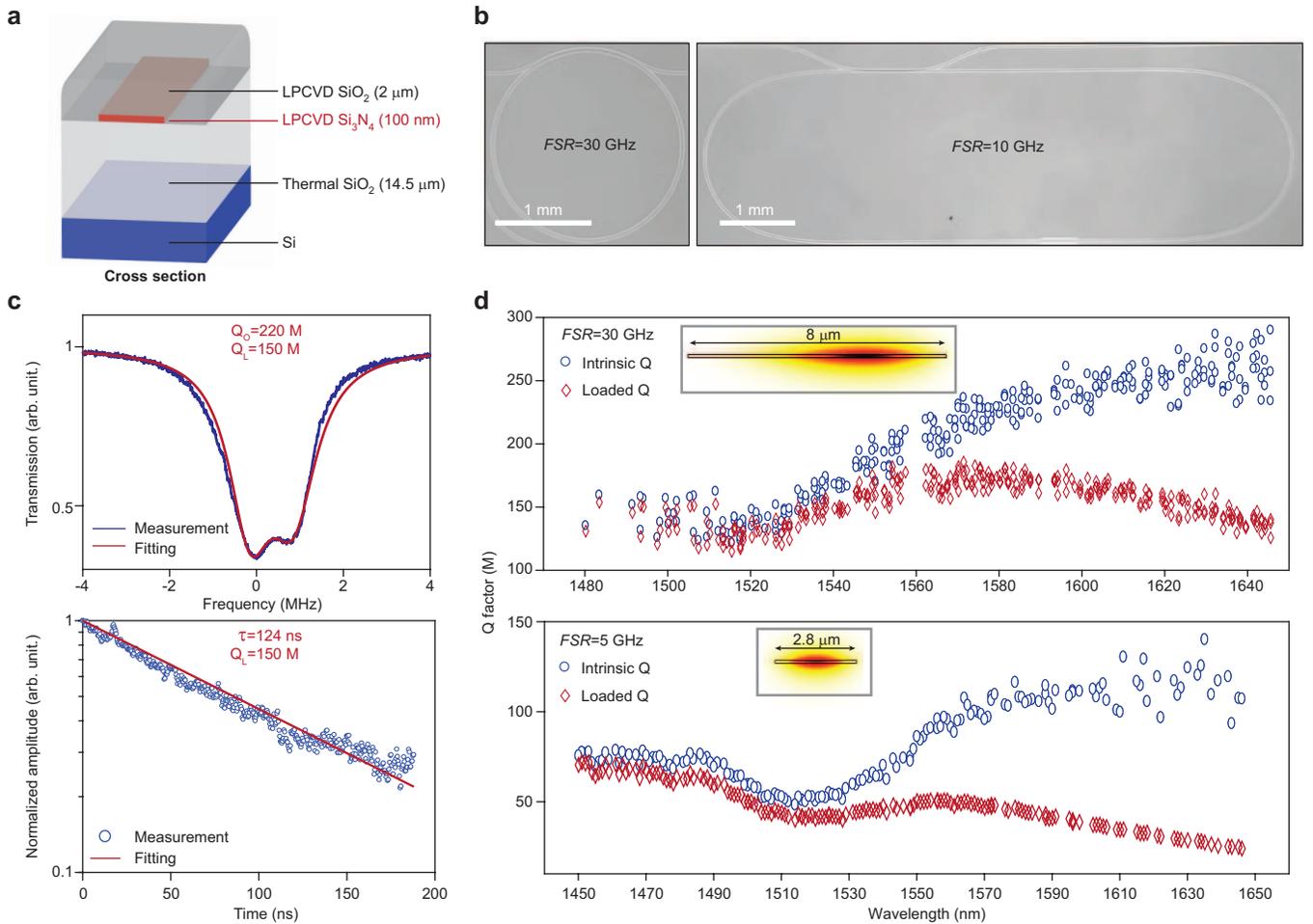}
\caption{{\bf Ultra-high-$Q$ Si$_3$N$_4$ microresonators.} {\bf a,} Cross sectional diagram of the ultra-low loss waveguide, consisting of Si$_3$N$_4$ as the core material, silica as the cladding, and silicon as the substrate (not to scale). {\bf b,} Top view of the Si$_3$N$_4$ microresonators with 30~GHz $FSR$ (ring, left panel) and 10~GHz $FSR$ (racetrack, right panel). {\bf c,} Transmission spectrum (upper panel) of a high-$Q$ mode at 1560~nm in a 30~GHz ring resonator. Interfacial and volumetric inhomogeneities induce Rayleigh scattering, causing resonances to appear as doublets due to coupling between counter-propagating modes. Intrinsic $Q$ of 220~M and loaded $Q$ of 150~M is extracted by fitting the asymmetric mode doublet. The ring-down trace of the mode (lower panel) shows 124~ns photon lifetime, corresponding to a 150~M loaded $Q$. {\bf d,} Measured intrinsic $Q$ factors plotted versus wavelength in a 30~GHz ring resonator with 8~$\mu$m wide Si$_3$N$_4$ core (upper panel) and a 5~GHz racetrack resonator with 2.8~$\mu$m wide Si$_3$N$_4$ core (lower panel). Insets: simulated optical mode profile.}
\label{figure1}
\end{figure*}

\medskip
\noindent{\bf Results}

\noindent{\bf CMOS-ready ultra-high-$Q$ microresonators}\par
\noindent The ultra-high $Q$ factor resonators use high-aspect-ratio Si$_3$N$_4$ waveguides as shown in Fig. \ref{figure1}a. The samples are fabricated in a high-volume CMOS foundry on 200~mm wafers following the process of Bauters et al. \cite{bauters2011planar}, but we increase the thickness of the Si$_3$N$_4$ core from 40~nm to 100~nm. Thicker Si$_3$N$_4$ enables a bending radius below 1~mm \cite{supplement}, allowing higher integration density than the centimeter-sized resonators demonstrated previously \cite{spencer2014integrated,gundavarapu2019sub,puckett2019silicon}. Furthermore, a top cladding thickness of 2 µm is sufficient, which obviates the need for complex chemical-mechanical polishing and bonding of additional thermal SiO$_2$ on top \cite{bauters2011planar, spencer2014integrated}.  Microresonators having three different  free spectral ranges ($FSR$) were fabricated. Those resonators having 30~GHz FSR were in a whispering-gallery-mode ring geometry while single-mode racetrack resonators with 5~GHz and 10~GHz $FSR$ were fabricated to reduce footprint (Fig.~\ref{figure1}b). All devices were fabricated on the same wafer. 

\begin{figure*}
\centering
\includegraphics[width=0.8\linewidth]{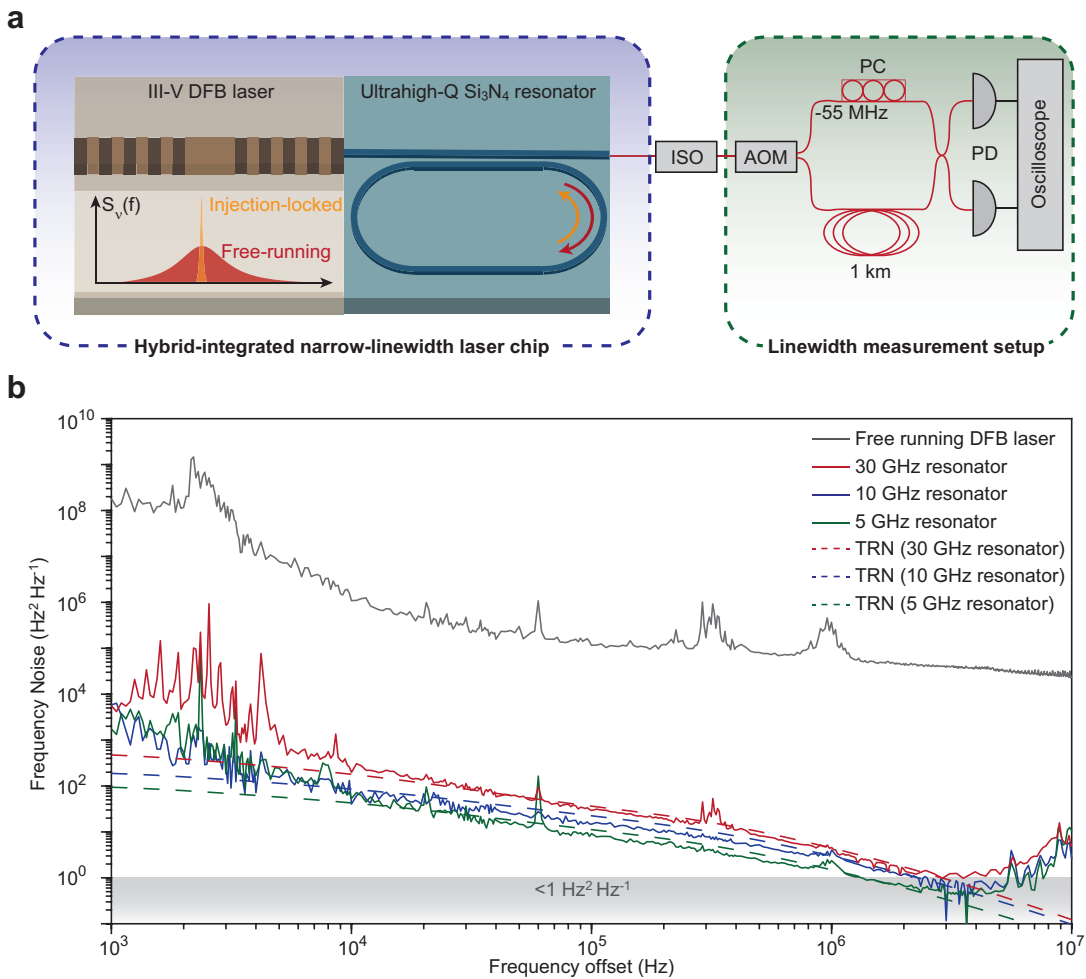}
\caption{{\bf Hybrid-integrated narrow-linewidth laser based on ultra-high-$Q$ Si$_3$N$_4$ microresonator.} {\bf a,} Schematic of the hybrid laser design (not to scale) and linewidth test setup. The red (yellow) arrow denotes the forward (backscattered) light field. ISO: optical isolator; AOM: acousto-optic modulator; PC: polarization controller; PD: photodetector. {\bf b,} Measurement of single-sideband frequency noise of the free-running and self-injection locked DFB laser. The white-frequency-noise levels are 1~Hz$^2$~Hz$^{-1}$, 0.8~Hz$^2$ Hz$^{-1}$,~0.5 Hz$^2$~Hz$^{-1}$ for resonators with 20~GHz, 10~GHz and 5~GHz $FSR$, respectively. The dashed lines give the simulated thermorefractive noise (TRN).}
\label{figure2}
\end{figure*}
 
Transmission spectra scans using a tunable external cavity laser (calibrated by a separate interferometer) were measured to study the resonator linewidth and to infer loaded, coupled and intrinsic optical $Q$ factors. Cavity ring down was also performed as a separate check of these $Q$ measurments. Spectra were observed to occur in doublets on account of both the ultra-high-Q and the presence of waveguide backscattering (Fig.~\ref{figure1}c) \cite{Kippenberg02}. By fitting the doublet line shape of the 30~GHz ring resonator, intrinsic $Q$ of 220~M and loaded $Q$ of 150~M are extracted at 1560 nm, which are further confirmed by measuring the ring-down trace of the resonance as shown in Fig.~\ref{figure1}c. The spectral dependences of $Q$-factors in ring- and racetrack-resonators (Fig.~\ref{figure1}d) provide insight into the origins of loss. A reduction in the value of $Q$ around 1510~nm is due to absorptive N-H bonds in the Si$_3$N$_4$ core. Beyond this wavelength, the intrinsic $Q$ factor increases monotonically versus wavelength, likely limited by Rayleigh scattering. The highest $Q$ factor is obtained using the 30~GHz FSR resonator (mean value of 260~M and standard deviation of 13.5~M over 34 modes) and observed in the 1630~nm to 1650~nm wavelength range. The overall lower $Q$ factor of the 5~GHz racetrack resonator suggests excess propagation loss in its single mode waveguides. This is possibly caused by higher scattering loss from increased modal overlap with the waveguide sidewall as compared to the whispering-gallery mode waveguide.

\medskip
\noindent{\bf Hertz-linewidth integrated laser}\par
\noindent The hybrid-integrated laser comprises a commercial DFB laser butt-coupled to the bus waveguide of the Si$_3$N$_4$ resonator chip (Fig.~\ref{figure2}a). The laser chip, which is mounted on a thermoelectric cooler to avoid long-term drift, is able to deliver power up to 30~mW at 1556~nm into the Si$_3$N$_4$ bus waveguide. Optical feedback is provided to the laser by backward Rayleigh scattering in the microresonator, which spontaneously aligns the laser frequency to the nearest resonator mode. As the phase accumulated in the feedback is critical to determining the stability of injection-locking \cite{kondratiev2017self,shen2020integrated,supplement}, we precisely control the feedback phase by adjusting the air gap between the chips. The laser output is taken through the bus waveguide of the microresonator, and directed to a self-heterodyne setup for linewidth characterization. Two photodetectors and a cross-correlation technique are used to improve detection sensitivity (see Methods).

\begin{figure*}[t!]
\centering
\includegraphics[width=\linewidth]{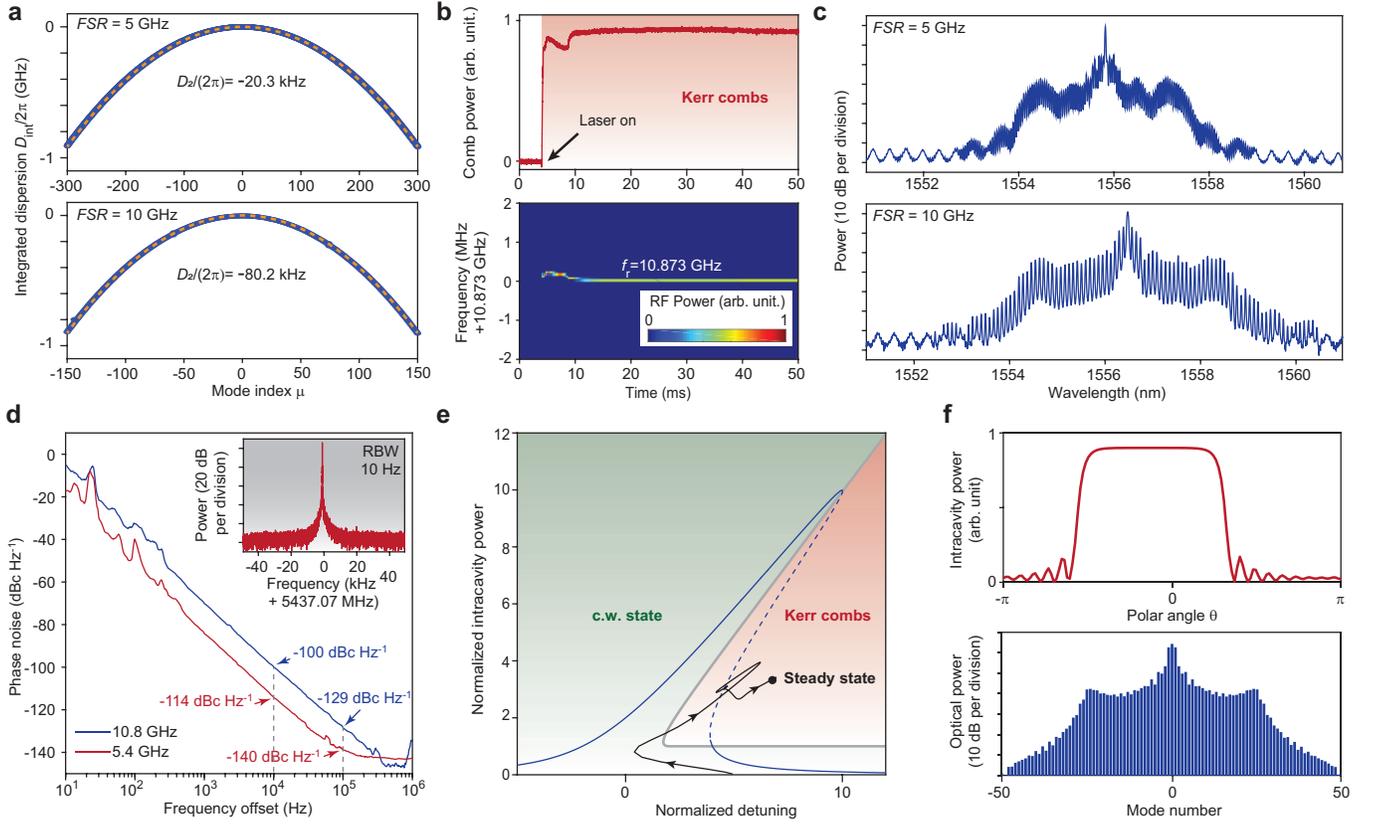}
\caption{{\bf Formation of mode-locked Kerr combs.} {\bf a,} Measured mode family dispersion is normal. The plot shows the integrated dispersion defined as $D_{\rm int}=\omega_\mu-\omega_o-D_1\mu$ where $\omega_\mu$ is the resonant frequency of a mode with index $\mu$ and $D_1$ is the $FSR$ at $\mu=0$. The wavelength of the central mode ($\mu=0$) is around 1550~nm. The dashed lines are parabolic fits ($D_{\rm int}= D_2 \mu^2 /2 $) with $D_2/2\pi$ equal to $-20.3$~kHz and $-80.2$~kHz corresponding to 5~GHz and 10~GHz $FSR$, respectively. Note: $D_2=-cD_1^2\beta_2/n_{\rm eff}$ where $\beta_2$ is the group velocity dispersion, $c$ the speed of light and $n_{\rm eff}$ the effective index of the mode. {\bf b,} Experimental comb power (upper panel) and detected comb repetition rate signal (lower panel) with laser turn-on indicated at 5~ms. {\bf c,} Measured optical spectra of mode-locked Kerr combs with 5~GHz (upper panel) and 10~GHz (lower panel) repetition rates. The background fringes are attributed to the DFB laser. {\bf d,} Single-sideband phase noise of dark soliton repetition rates. Dark solitons with repetition rate 10.8~GHz and 5.4~GHz are characterized. Inset: electrical beatnote showing 5.4~GHz repetition rate. {\bf e,} Phase diagram of microresonator pumped by an isolated laser. The backscattering is assumed weak enough to not cause mode-splittings. The detuning is normalized to one half of microresonator linewidth, while the intracavity power is normalized to parametric oscillation threshold. Green and red shaded areas indicate regimes corresponding to the c.w. state and Kerr combs. The blue curve is the c.w. intracavity power, where stable (unstable) branches are indicated by solid (dashed) lines.  Simulated evolution of the unisolated laser is plotted as the solid black curve, and it converges to the steady state as marked by the black dot. The initial condition is set within the self-injection locking bandwidth \cite{shen2020integrated}, while feedback phase is set to 0. {\bf f,} Simulated intracavity field (upper panel) and optical spectrum (lower panel) of the unisolated laser steady state in panel e.}
\label{figure3}
\end{figure*}

The frequency noise spectra of the self-injection locked laser system using the 30~GHz ring, and the 10~GHz and 5~GHz racetrack resonators (respective intrinsic $Q$ factors of 250~M, 56~M and 100~M) are compared in Fig. \ref{figure2}b. The ultra-high-Q factors enable the frequency noise of the free-running DFB laser to, in principle, be suppressed by up to 80~dB (see Methods). In practice, however, the noise suppression over a broad range of offset frequencies (10~kHz to 2~MHz) is limited to 50~dB by the presence of thermorefractive noise \cite{kondratiev2018thermorefractive,huang2019thermorefractive} in the microresonator. Consistent with theory, microresonators with larger mode volume, i.e. smaller $FSR$, experience a lower thermorefractive fluctuation and exhibit reduced frequency noise (Fig. \ref{figure2}b). At low frequency offset (below 10~kHz), frequency noise is primarily limited by temperature drift and coupling stability between chips. This can be suppressed by improved device packaging. At high offset frequencies (above 5~MHz), frequency noise rises with the square of offset frequency, as the maximum noise suppression bandwidth of injection locking is limited to the bandwidth of the resonator \cite{supplement, hjelme1991semiconductor}. Thus, minimum frequency noise below 1~Hz$^2$~Hz$^{-1}$ is observed at about 5~MHz offset frequency, where the contributions of rising laser noise and falling thermorefractive noise are approximately equal. To achieve an ultra-low white frequency noise floor at high offset frequencies, the laser output may be taken from a resonator featuring a drop-port. The drop port would provide low-pass filtering action and is studied further in the Supplement \cite{supplement}.

\begin{figure}[t!]
\centering
\includegraphics[width=\linewidth]{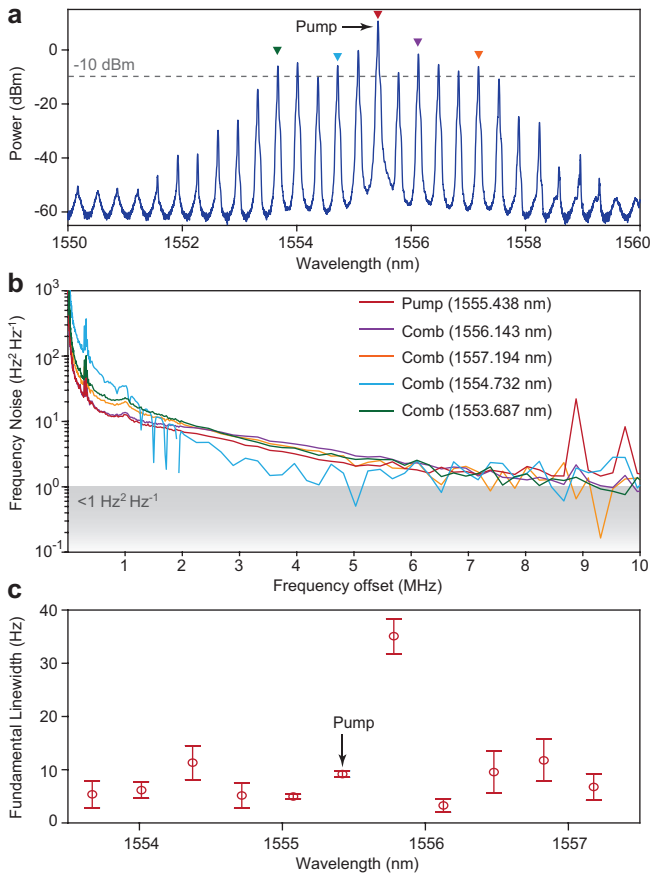}
\caption{{\bf Coherence of integrated mode-locked Kerr combs.} {\bf a,} Optical spectrum of a mode-locked comb with 43.2~GHz repetition rate generated in a microresonator with 10.8~GHz $FSR$. {\bf b,} Single-sideband optical frequency noise of the pump and comb lines as indicated in panel a, selected using a tunable fiber-Bragg-grating (FBG) filter. {\bf c,} Wavelength dependence of white frequency noise linewidth of comb lines in panel a.}
\label{figure4}
\end{figure}

\medskip
\noindent{\bf Mode-locked Kerr comb}\par
\noindent The ultra-high Q of the microresonators enables strong resonant build-up of the circulating intensity, providing access to nonlinear optical phenomena at low input power levels \cite{vahala2003optical}. As an example, optical frequency combs have been realized in continuously pumped high-$Q$ optical microresonators due to the Kerr nonlinearity and they are finding a wide range of applications \cite{Kippenberg2018}. To explore the nonlinear operating regime of the hybrid-integrated laser in pursuit of highly-coherent Kerr combs, the mode dispersion of racetrack resonators with 5~GHz and 10~GHz $FSR$ was characterized. Their mode families are measured to have normal dispersion across the telecommunication C-band (Fig. \ref{figure3}a). Also, the dispersion curves exhibit no avoided-mode-crossings, which is consistent with the single-mode nature of the waveguides. As distinct from microresonators with anomalous dispersion wherein bright soliton pulses are readily generated, comb formation is forbidden in microresonators with normal dispersion, unless avoided-mode-crossings are introduced to alter mode family dispersion so as to allow formation of dark solitons \cite{xue2015mode}. Surprisingly, however, it was nonetheless possible to readily form coherent combs in these devices without either of the aforementioned conditions being satisfied. 

Indeed, deterministic, turnkey comb formation was experimentally observed when the DFB laser was switched-on to a preset driving current (see Fig.~\ref{figure3}b). A clean and stable beatnote of the comb is established 5~ms after turning on the laser, indicating that mode-locking has been achieved (see Fig.~\ref{figure3}b). Plotted in Fig.~\ref{figure3}c are optical spectra of the mode-locked Kerr combs in resonators with 5~GHz and 10~GHz $FSR$, where the typical spectral shape of dark soliton pulses is observed \cite{liang2014generation,xue2015mode,lobanov2015frequency,Kippenberg2018}. The stability of mode-locking is characterized by measurement of the comb beat note phase noise (Fig.~\ref{figure3}d). For Kerr combs with 10.8 (5.4) GHz $FSR$, the phase noise reaches -100 (-114) dBc~Hz$^{-1}$ at 10~kHz and -129 (-140) dBc~Hz$^{-1}$ at 100~kHz offset frequencies. We note that in order to suppress noise at high-offset frequencies, the pump is excluded in the photodetection using a fiber Bragg grating filter, as suggested by previous works \cite{liang2015high}.

This unexpected result is studied theoretically in the Supplement. Here, results from that study are briefly summarized. A phase diagram of the microcomb system is given in Fig. \ref{figure3}e, and separates resonator operation into continuous-wave (c.w.) and Kerr comb regimes based on the viability of parametric oscillation \cite{godey2014stability}. The intracavity power exhibits a typical bi-stable behavior as a function of cavity-pump frequency detuning when pumped by a laser with optical isolation \cite{Kippenberg2018}. In contrast, a recent study shows that the feedback from a nonlinear microresonator to a non-isolated laser creates an operating point for the compound laser-resonator system in the middle branch \cite{shen2020integrated}. The operating point is induced through a combination of self- and cross-phase modulation, and is associated with turnkey operation of soliton combs operating under conditions of anomalous dispersion \cite{shen2020integrated}.  Here, we have validated through simulation that the same operating point allows access to dark solitons (normal dispersion) without the requirement for extra dispersion engineering provided by avoided mode crossings.  The black curve in Fig.~\ref{figure3}e gives the dynamics of the compound laser-resonator system when initialized at a point that is within the locking bandwidth of the system. It converges to a steady state located in the Kerr comb regime. The spectral and temporal profile of the steady state solutions show that flat-top pulses are formed in the microresonator with normal dispersion (Fig.~\ref{figure3}f). The possible presence of dark soliton formation in microresonators pumped by a self-injection locked laser has been observed, but has not yet been clarified previously \cite{liang2014generation,lobanov2015frequency}.

The combs generated in these devices exhibit several important properties. In Fig.~\ref{figure4}a, the spectrum of a 43.2~GHz repetition rate comb is presented. Curiously, this spectrum was generated in a microresonator having a 10.8~GHz $FSR$. The appearance of rates that are different from the $FSR$ rate has been observed for dark solitons \cite{xue2015mode}. This line spacing is compatible with DWDM channel spacings and 10 comb teeth feature on-chip optical power over -10~dBm, which is a per channel power that is readily usable in DWDM communication systems \cite{fulop2018high}. However, most significant, is that the white-frequency-noise-level floor for each of these optical lines (Fig. \ref{figure4}b) is measured to be on the order of 1~Hz$^2$~Hz$^{-1}$. We note that these spectra are truly white, i.e.,  not rising for higher offset as discussed above for the laser source. The corresponding fundamental linewidths of the comb teeth are plotted in Fig. \ref{figure4}c. One of the lines exhibits degraded linewidth of approximately 30~Hz, which is suspected to be due to its coincidence with a sub-lasing-threshold side-mode of the DFB laser. Notably, certain comb teeth are quieter than the pump due to the filtering of pump noise by the ultra-high-$Q$ modes. These results represent a two order-of-magnitude improvement as compared to previously demonstrated integrated microcombs \cite{stern2018battery,raja2019electrically,shen2020integrated}.

\begin{table}
\renewcommand{\figurename}{\textbf{Table.}}
  \includegraphics[width=\linewidth]{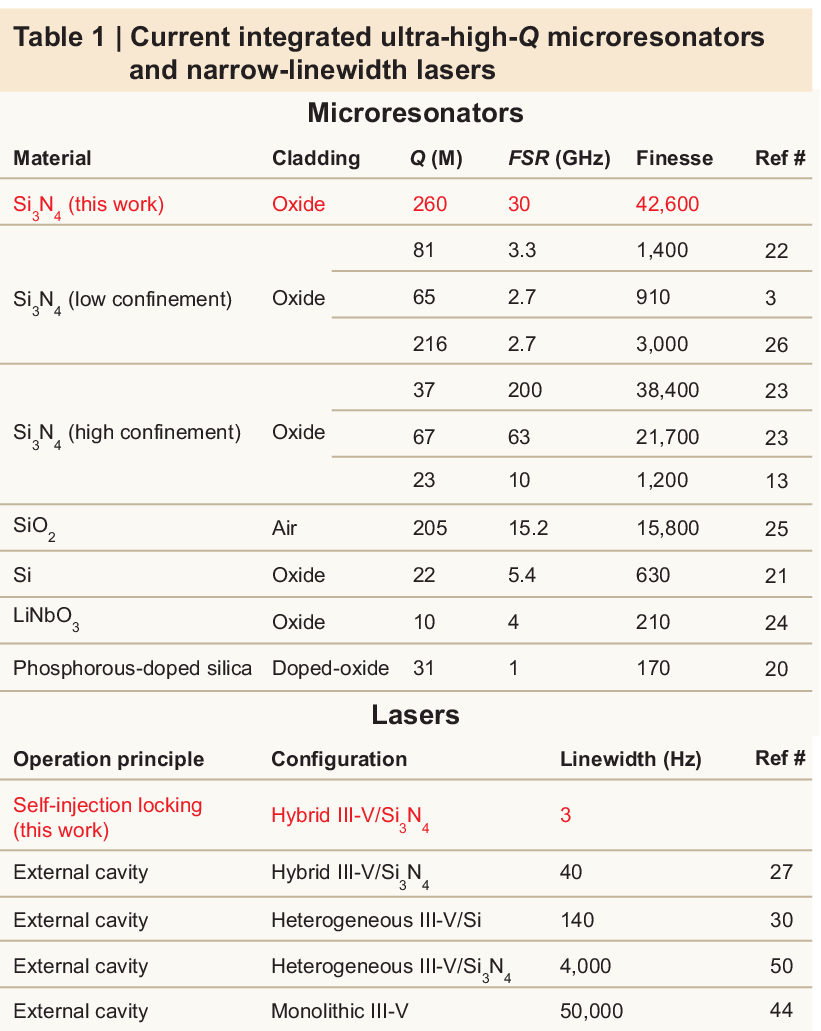}
  \caption{{\bf Upper: Best-to-date integrated ultra-high-$Q$ ($>$~10~M) microresonators with integrated waveguides. Lower: Best-to-date integrated narrow-linewidth lasers.} 
  }
  \label{Table1}
\end{table}

\medskip
\noindent{\bf Performance Comparison}\par
\begin{figure}
\centering
\includegraphics{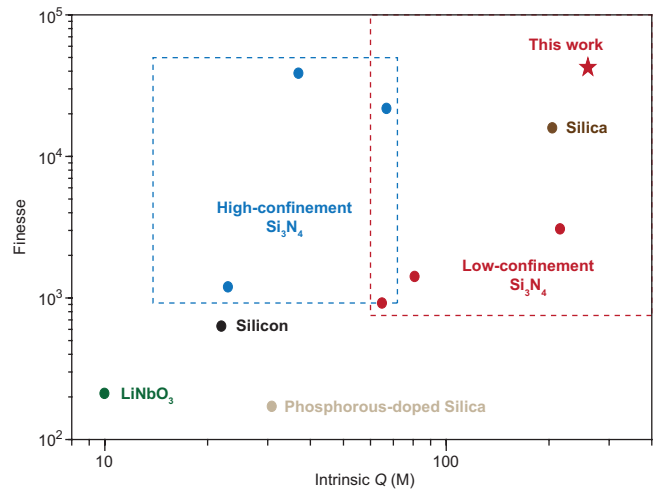}
\caption{{\bf Comparison of finesse and intrinsic $Q$ factors of state-of-the-art integrated microresonators.}}
\label{figure5}
\end{figure}

\noindent For devices with both integrated waveguide coupler and resonator, a few platforms have emerged as able to provide ultra-high $Q$ ($Q>10$ M). In silica ridge resonators, a $Q$ factor of 205~M has been demonstrated \cite{yang2018bridging}, while in low-confinement silicon nitride, a $Q$ factor of 216~M has been demonstrated \cite{puckett2019silicon}. However, these platforms pose challenges to photonic integration with large scale and high density, e.g. the use of suspended structures \cite{yang2018bridging} or the requirement for centimeter-level bending radius \cite{puckett2019silicon}. While these limitations are not present in high-confinement silicon nitride resonators, the highest demonstrated $Q$ factor is lower, 67~M \cite{ji2017ultra}. In Table \ref{Table1}, we list key figures of merit for integrated microresonators with ultra-high-$Q$ factors. In addition to record-high Q factor, owing to their compact footprint, the current resonators stand out among ultra-high Q resonators for having the highest finesse as well. Fig.~\ref{figure5} provides a comparison as a plot of the $Q$ and finesse of the current work with the state-of-the-art.

We further compare the current hybrid-integrated laser linewidth to state-of-the-art results in Table~\ref{Table1}. The Lorentzian linewidth of monolithic III-V lasers is generally limited to the 100~kHz to 1~MHz range by passive waveguide losses well above 1~dB~cm$^{-1}$, with best demonstrated linewidth below 100~kHz \cite{larson2015narrow}. Phase and amplitude noise scale according to the square of cavity losses \cite{hjelme1991semiconductor,kondratiev2017self}. Thus, hybrid integration, where the active III-V and passive photonic chips are assembled post-fabrication, and heterogeneous integration \cite{komljenovic2016heterogeneous}, where III-V material is directly bonded to the passive chip during fabrication, have emerged as primary technologies to create narrow-linewidth integrated lasers. As shown in Table~\ref{Table1}, hybrid and heterogeneous integration can produce fundamental linewidth well below 1~kHz. In this work, fundamental frequency noise is suppressed to 0.5~Hz$^2$~Hz$^{-1}$, or equivalently, a 3~Hz linewidth, which is more than an order of magnitude improvement over the best results to date \cite{fan2019ultra}. 

\medskip
\noindent{\bf Discussion}

\noindent As single-frequency or mode-locked lasers, these hybrid-integrated devices are readily applicable to many coherent optical systems. For example, while laboratory communication experiments pursuing spectral efficiency approaching 20~bit~s$^{-1}$~Hz$^{-1}$ rely on high performance single-frequency fiber lasers \cite{olsson2018probabilistically}, narrow-linewidth integrated photonic comb lasers could accelerate the adoption of similar schemes in practical data-center and metro links \cite{marin2017microresonator,fulop2018high,raja2019electrically,stern2018battery,corcoran2020ultra,shen2020integrated}. Microwave photonics \cite{Li2013,liang2015high,hao2018toward,marpaung2019integrated,liu2020photonic}, atomic clocks \cite{ludlow2015optical,newman2019architecture}, and quantum information \cite{tanzilli2012genesis} will also benefit greatly from the reduced size, weight, power and cost provided by the combination of ultra-high $Q$ and photonic integration.

Many improvements beyond the results presented here are feasible. We infer propagation loss of 0.1~dB~m$^{-1}$, however, lower loss of 0.045~dB~m$^{-1}$ is feasible in thinner cores \cite{bauters2011planar}, suggesting that the limits of $Q$ for this platform have not been fully explored. Spiral resonators with increased modal volume can suppress low-offset frequency noise induced by thermodynamic fluctuations \cite{Lee2014spiral}. Finally, heterogeneous integration of III-V lasers and ultra-high-$Q$ microresonators may eventually unite the device onto a single chip \cite{tran2019tutorial,xiang2020narrow,komljenovic2016heterogeneous}, leading to scalable production with high yield using foundry-based technologies.

\bibliography{scibib}

\clearpage
\noindent {\bf\Large Methods} \\

\noindent{\bf Experimental details}\\
The $Q$ is obtained by frequency-down-scanning a external-cavity-diode-laser (ECDL) across a mode, with frequency calibrated using a Mach-Zehnder interferometer (MZI). Measured transmission spectra at various wavelengths are shown in the Supplement \cite{supplement}. Similarly, the mode family dispersion is extracted from the broadband transmission spectrum of the resonator, with calibration provided by the MZI as well.

The laser switch-on test is performed by rapidly modulating its driving current with square wave functions. The real-time evolution of comb repetition rate is obtained by down-mixing the photodetected beatnote with a local microwave oscillator. The trace, which is recorded using a high-speed oscilloscope, is Fourier transformed to give the spectrograph. Multiple turnkey tests are shown with 100 percent success rate (see Supplement \cite{supplement}). The phase noise of comb repetition rates is characterized using a Rohde \& Schwarz phase noise analyzer.

\medskip

\noindent{\bf Laser linewidth measurement}\\
The noise in the photodetection, e.g., shot noise, thermal noise and dark current, limit the sensitivity of self-heterodyne method especially at high-offset frequencies. To overcome such limit, we use two photodetectors to measure the self-heterodyne signals simultaneously. The instantaneous frequency is extracted using Hilbert transformation, and their cross correlation $C_\nu (f)$ is given by
\begin{equation}
\begin{split}
C_\nu(f) &=2\left[1-\left(1-\tau_0\mathrm{BW}\right)^+\cos(2\pi f \tau_0)\right]S_\nu(f)\\
&-\frac{1}{2}\left[1+\left(1-\tau_0\mathrm{BW}\right)^+\cos(2\pi f \tau_0)\right]\\
&\times f^2 (S_I(f_c+f)+S_I(f_c-f))\\
\end{split}
\end{equation}
where $S_\nu(f)$ and $S_I(f)$ are the single-sideband power spectral density of frequency and relative intensity noise (RIN) of the laser, $\tau_0$ the delay between the two arms, and $x^+=\max(0,x)$ the ramp function. The resolution bandwidth of the cross-correlator $\mathrm{BW}$ is set as 20 kHz so that $\tau_0 \mathrm{BW}\ll 1$. To reduce the contribution of RIN as well as enhance the detection sensitivity of frequency noise, at high-offset frequencies ($f>2$ MHz) we only select the data where $\cos(2\pi f \tau_0)\approx-1$. The enhancement of sensitivity equals $\sqrt{{\rm BW}*T}$ with $T$ the recording time. In this measurement $T$ is set 200 ms, corresponding to 18 dB enhancement of sensitivity.

\medskip
\noindent{\bf Thermorefractive noise}\\
Constant heat exchange between the microresonator and its ambient results in thermodynamic fluctuations, which could induce changes in the refractive index through thermo-optic effect, giving rise to thermorefractive noise of the resonant frequencies \cite{kondratiev2018thermorefractive,huang2019thermorefractive}. The variance of the thermorefractive noise (TRN) is given by
\begin{equation}
    <\delta\omega_c^2>=\frac{n_{\rm T}^2\omega_c^2}{n_{\rm eff}^2}\frac{k_{\rm B}T^2}{\rho C V}.
\end{equation}
where $n_{\rm T}$ is the thermo-optic coefficient, $\omega_c$ the resonant frequency, $n_{\rm eff}$ the effective index of the mode, $k_{\rm B}$ the Boltzmann's constant, $T$ the temperature of the heat bath, $\rho$ the density, $C$ the specific heat and $V$ the volume. Owing to their larger mode volumes, the low-confinement resonators in this work feature notably smaller TRN than those of high-confinement resonators \cite{huang2019thermorefractive}. The spectral density of the TRN is computed using finite-element-method (FEM) based on fluctuation-dissipation theorem \cite{kondratiev2017self}, as plotted in Fig. \ref{figure2}b in the maintext.

\medskip
\noindent{\bf Linewidth-reduction factor}\\
The amount of linewidth-reduction in self-injection locked laser depends on the spectral response and power of the backscattered field, which has been derived in the supplement based on a complete theory involving both laser and microresonator dynamics \cite{kondratiev2017self}. We introduce the coupling between the clockwise and counterclockwise field in the microresonator, $\beta$, which is normalized to one half of the cavity linewidth. In the case of weak backscattering ($\beta\ll 1$), i.e., the mode remains as a singlet, the laser linewidth can be reduced by
\begin{equation}
    \alpha\approx 64(1+\alpha_g^2)T^2\eta^2|\beta|^2\frac{Q_{\rm R}^2}{Q_{\rm d}^2},
\end{equation}
where $Q_{\rm R}$ and $Q_{\rm d}$ stand for the $Q$ of the microresonator and the laser diode, respectively. $\eta=Q_{\rm R}/Q_{\rm e}$ is the microresonator loading factor with $Q_{\rm e}$ being the coupling $Q$ between the bus waveguide and the resonator. $T$ denotes the power insertion loss between the facets of the laser and the bus waveguide, while $\alpha_g$ is the amplitude-phase coupling coefficient of the laser. In the presence of a strong backscattered field ($\beta\gg 1$), i.e., the mode splits into doublets, the linewidth-reduction factor is saturated as
\begin{equation}
    \alpha\approx 4(1+\alpha_g^2)T^2\eta^2\frac{Q_{\rm R}^2}{Q_{\rm d}^2},
\end{equation}
which is independent of the backscattering coefficient. Typical values of these parameters in our systems are: $\alpha_g=2.5$, $T=-6$ dB, $\eta=0.5$, $Q_{\rm d}=10^4$. For mode featuring loaded $Q$ of 50~M and split resonances, the maximum estimated noise reduction factor is around 70~dB, which is 20 dB higher than the noise suppression achieved in experiment. In the experiment, the locking point is intentionally offset from the exact resonance by adjusting the feedback phase to avoid nonlinearity.

\medskip
\noindent{\bf Phase diagram}\\
The phase diagram presented in Fig. \ref{figure3}b of the maintext is a powerful tool to interpret how self-injection locking can deterministically lead to mode-locked Kerr comb formation. Assuming homogeneous intracavity field, the parametric gain of the $\pm l_{\rm th}$ modes relative to the pump is given by \cite{godey2014stability}
\begin{equation}
    \Gamma(\pm l)={\rm Re} \left\{-1+\sqrt{\rho^2-(\Delta-2\rho+d_2 l^2)^2} \right\},
\end{equation}
where $\rho^2$ is the intracavity power normalized to the parametric oscillation threshold \cite{Kippenberg2018}, $\kappa$ represents the modal linewidth, $\Delta=2\delta\omega/\kappa$ the normalized detuning, $\delta\omega$ the pump-cavity detuning, and $d_2=D_2/\kappa$ the normalized dispersion. To initiate parametric oscillation, $\Gamma(\pm l)>0$ is required. At the minimal value of $l^2=1$, the regime corresponding to Kerr comb is given by
\begin{equation}
    \Delta>2\rho+d_2-\sqrt{\rho^2-1}.
\end{equation}

\medskip
\noindent {\bf\large Data availability} \\
\noindent All data generated or analysed during this study are available within the paper and its supplementary materials. Further source data will be made available on reasonable request.\\

\noindent {\bf\large Code availability} \\
\noindent The analysis codes will be made available on reasonable request.\\

\noindent {\bf Acknowledgments} The authors gratefully acknowledge the Defense Advanced Research Projects Agency (DARPA) under DODOS (HR0011-15-C-055) programs and Anello Photonics.\\
\\
\noindent{\bf Author contributions} Experiments were conceived by W.J., Q.-F.Y., L.C., B.S. and H.W. Devices were designed by W.J., and A.F. Measurements were performed by W.J., Q.-F.Y., L.C, B.S., H.W. with assistance from M.A.L and L.W. Analysis of results was conducted by W.J., Q.-F.Y. and H. W. The project is coordinated by Q.-F.Y. and L.C. under the supervision from J.B., K.V. and M.P. All authors participated in writing the manuscript.\\
\\
\noindent{\bf Competing interests} The authors declare no competing financial interests.\\
\\
\noindent{\bf Additional information} \\
\noindent{\bf Supplementary information} is available for this paper.\\
\noindent{\bf Correspondence and requests for materials} should be addressed to K.V. and J.B.

\end{document}


\title{Supplementary Information to Hertz-linewidth semiconductor lasers using CMOS-ready ultra-high-Q microresonators}
\author{
Warren Jin$^{1,\ast}$, Qi-Fan Yang$^{2,\ast}$, Lin Chang$^{1,\ast}$, Boqiang Shen$^{2,\ast}$, Heming Wang$^{2,\ast}$,\\Mark A. Leal$^{1}$, Lue Wu$^{2}$, Avi Feshali$^{3}$, Mario Paniccia$^{3}$, Kerry J. Vahala$^{2,\dagger}$, and John E. Bowers$^{1,\dagger}$\\
$^1$ECE Department, University of California Santa Barbara, Santa Barbara, CA 93106, USA\\
$^2$T. J. Watson Laboratory of Applied Physics, California Institute of Technology, Pasadena, CA 91125, USA\\
$^3$Anello Photonics, Santa Clara, CA\\
$^\ast$These authors contributed equally to this work.\\
$^\dagger$Corresponding authors: vahala@caltech.edu, bowers@ece.ucsb.edu\\
}

\maketitle
\section{Device design}

\begin{figure}
\centering
\includegraphics{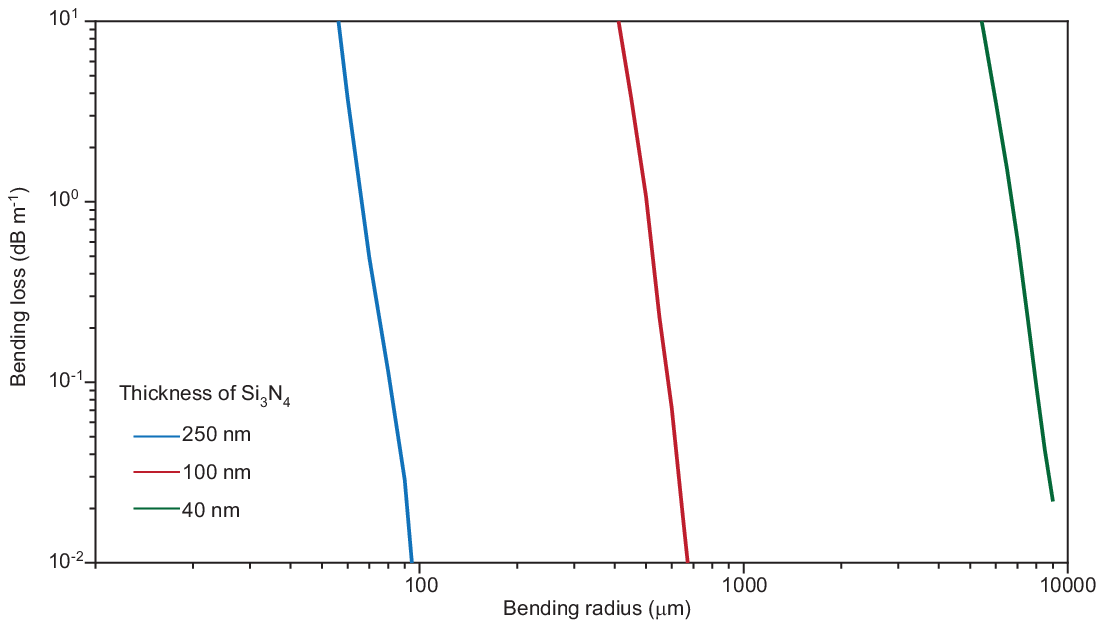}
\caption{{\bf Bending loss versus radius.} Simulated waveguide bending loss of single mode waveguides at three thicknesses of Si$_3$N$_4$ core.}
\label{figureS1}
\end{figure}

Among the design parameters involved in achieving ultra-high $Q$ factors, the choice of waveguide thickness is critical, as thinner cores will minimize the impact of scattering losses through lower modal confinement, at the expense of larger bending radius and thus lower integration density. In Fig.~\ref{figureS1}, we compare simulated bending loss in waveguides of 250~nm, 100~nm, and 40~nm thickness and 1.4~$\mu$m, 2.8~$\mu$m, and 5.6~$\mu$m width, respectively, where width is chosen to guarantee single-mode propagation within the C~band. Although the lowest propagation loss has been demonstrated in 40~nm thick cores \cite{bauters2011planar}, a radius of 1~cm is required to guarantee bending loss below 0.01 dB/m, which limits the integration density to a single resonator per die for typical lithography stepper reticles of 2~cm by 2~cm dimension. On the other hand, 250~nm thick cores allow for highly compact resonators, with just 100~$\mu$m bend radius. In this work, we focus on achieving ultra-low loss while maintaining moderate integration density. Thus, we select 100~nm as the optimal core thickness to allow bend radius as low as 700~$\mu$m, permitting as many as 100 resonators to be integrated on a single die while still suppressing the contribution of the waveguide sidewall roughness to scattering loss.  For the resonators designed in this work, the minimum bend radius is chosen as 1~mm, ensuring that bend losses are entirely negligible.

\begin{figure}
\centering
\includegraphics[width=\linewidth]{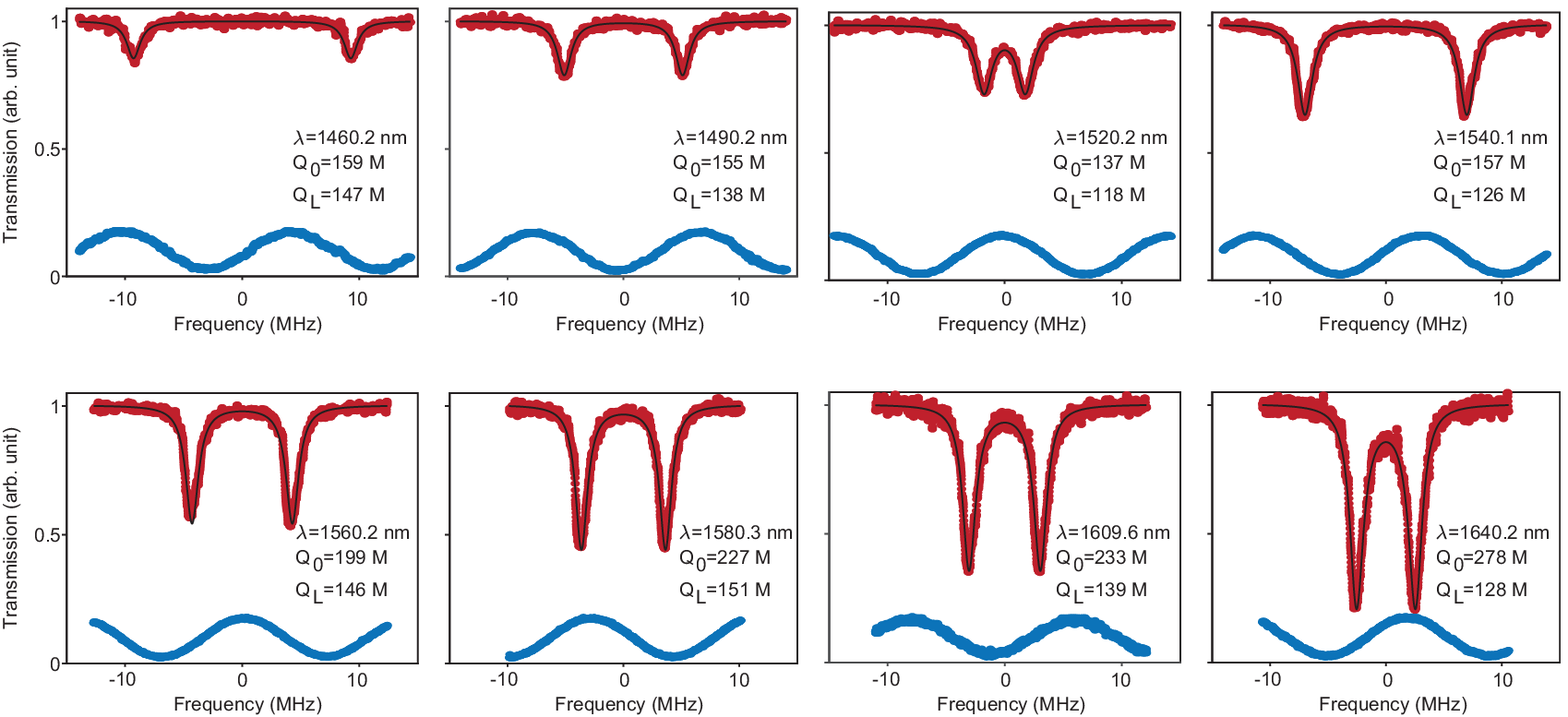}
\caption{{\bf $Q$ factor measurement.} Transmission of high-$Q$ modes at various wavelength in a 30~GHz resonator for determination of intrinsic and loaded $Q$ factors (red). Fitted doublet lineshape is overlaid. The tuning rate of the laser is simultaneously calibrated by the transmission of an MZI with 14.3~MHz $FSR$, also pictured (blue). }
\label{figureSQ}
\end{figure}

To achieve the highest $Q$ factor, we use multi-mode ring resonators to further minimize the contribution of sidewall scattering. For 1~mm bend radius, the fundamental mode propagation constant is insensitive to waveguide width for width beyond 8~$\mu$m. Hence, we choose a width of 8~$\mu$m, which produces a whispering-gallery mode and free spectral range ($FSR$) of 30~GHz. To ensure selective, low-loss coupling to the fundamental mode, a pulley coupler \cite{spencer2014integrated} is used. The coupler waveguide width is chosen to precisely phase-match the propagation constant of the coupler waveguide mode with that of the ring waveguide for a 5~$\mu$m width space between the waveguides. The length of the coupler is approximately 1~mm long, and is designed to be under-coupled to allow accurate measurement of the intrinsic $Q$ factor. Measured resonator transmission spectra at various wavelength are provided in Fig.~\ref{figureSQ}.

To achieve larger mode volume and smaller $FSR$ while retaining reasonable footprint and ultra-high $Q$, racetrack resonators with 5~GHz and 10~GHz $FSR$ are used. Racetrack resonators were designed for critical coupling at 1550~nm to ensure efficient delivery of power to the resonator. To convert the mode between straight and bending waveguides, a sine curve evaluated between 0 and $\pi/2$ is used as an adiabatic transition between the two. However, the transition between straight and bending waveguides still leads to residual conversion of the fundamental mode to unguided radiation modes, which causes excess loss. This is indicated by the lower intrinsic $Q$ factor of the 10~GHz racetrack resonator as compared to the 5~GHz resonator of 56~M and 100~M at 1560~nm, respectively. Imperfect adiabaticity can also cause coupling to higher-order waveguide modes. When those higher order modes are antiresonant, this coupling is suppressed, however strong coupling can occur at mode crossings. As the loss in higher-order modes typically exceeds that of the fundamental, the $Q$ at a mode crossing can be further degraded. Furthermore, the location of these mode crossings can be highly variable due to fabrication variability. Thus, to simplify operation of the device, we use single-mode waveguides within all racetrack resonators, at the expense of lower $Q$ factor. Nonetheless, the $Q$ factor is still high enough that the frequency noise of the self injection-locked laser is limited by thermorefractive noise, so the impact of reduced $Q$ factor is negligible.

\section{Theory: self-injection locking}

In this section we apply the theory of self-injection locking to study the linewidth-reduction factor when the laser is stabilized to the microresonator. The hybrid-laser system can be split into three parts: the laser field $A_{\rm L}$, the forward optical field $A_{\rm F}$ and the backscattered field $A_{\rm B}$ in the microresonator. The complete set of equations of motion are \cite{kondratiev2017self,shen2020integrated}:
\begin{equation}
\begin{split}
    &\frac{\partial A_{\rm F}}{\partial t}=-(\frac{\kappa}{2}+i\delta\omega)A_{\rm F}+i\frac{D_2}{2}\frac{\partial^2 A_{\rm F}}{\partial \theta^2}+ig(2\overline{|A_{\rm B}|^2}+|A_{\rm F}|^2)A_{\rm F}+i\beta\frac{\kappa}{2}\overline{A_{\rm B}}-\sqrt{T\kappa_{\rm R}\kappa_{\rm L}}e^{i\phi_{\rm B}}A_{\rm L},\\
    &\frac{\partial A_{\rm B}}{\partial t}=-(\frac{\kappa}{2}+i\delta\omega)A_{\rm B}+i\frac{D_2}{2}\frac{\partial^2 A_{\rm B}}{\partial \theta^2}+ig(2\overline{|A_{\rm F}|^2}+|A_{\rm B}|^2)A_{\rm B}+i\beta\frac{\kappa}{2}\overline{A_{\rm F}},\\
    &\frac{dA_{\rm L}}{dt}=i(\delta\omega_{\rm L}-\delta\omega)A_{\rm L}-\frac{\gamma_{\rm L}}{2}A_{\rm L}+\frac{g_{\rm L}}{2}(1+i\alpha_g)A_{\rm L}-\sqrt{T\kappa_{\rm R}\kappa_{\rm L}}e^{i\phi_{\rm B}}A_{\rm B}.
    \end{split}
\end{equation}
Here $\kappa$ and $\gamma_{\rm L}$ are the damping rates of the microresonator mode and the laser, respectively. $\delta\omega=\omega_c-\omega_o$ is the detuning of the cold-cavity resonant frequency $\omega_c$ relative to the injection-locked laser frequency $\omega_o$. Similarly, $\delta\omega_{\rm L}=\omega_c-\omega_{\rm L}$ with $\omega_{\rm L}$ being the free-running, cold laser frequency. The nonlinear coupling coefficient $g=\omega_o c n_2/n_{\rm eff}^2 V$ with $c$ the speed of light, $n_2$ the Kerr coefficient, $n_{\rm eff}$ the effective index and $V$ the effective mode volume. $\theta$ is the polar angle of the resonator in rotational frame, and $D_2$ is the group velocity dispersion. $\beta$ stands for the dimensionless backscattering coefficient normalized to the mode linewidth $\kappa/2$, which is chosen as a positive real number for simplicity. $\kappa_{\rm R}$ and $\kappa_{\rm L}$ denote the external coupling rate of the resonator and the laser facet, respectively. $T$ is the energy coupling efficiency between the facets of the laser and the Si$_3$N$_4$ waveguide, and $\phi_{\rm B}$ is the propagation phase delay between the resonator and the laser. $g_{\rm L}$ is the intensity dependent gain of the laser, and $\alpha_g$ is the amplitude-phase coupling coefficient. The average intracavity intensity $\overline{|A_{\rm F,B}|^2}=\int_{0}^{2\pi}|A_{\rm F,B}|^2 d\theta/2\pi$, while the average amplitude $\overline{A_{\rm F,B}}=\int_{0}^{2\pi}A_{\rm F,B} d\theta/2\pi$.

Introducing the normalized amplitude $\rho_{\rm F,B,L}=A_{\rm F,B,L}/\sqrt{E_{\rm th}}$ with $E_{\rm th}=\kappa/(2g)$ the intracavity parametric oscillation threshold, the equations of motion can be normalized to a dimensionless form
\begin{equation}
\begin{split}
    &\frac{\partial\rho_{\rm F}}{\partial\tau}=-(1+i\Delta)\rho_{\rm F}+id_2\frac{\partial^2 \rho_{\rm F}}{\partial \theta^2}+i(2\overline{|\rho_B|^2}+|\rho_F|^2)\rho_{\rm F}+i\beta \overline{\rho_{\rm B}}+F,\\
    &\frac{\partial \rho_{\rm B}}{\partial\tau}=-(1+i\Delta)\rho_{\rm B}+id_2\frac{\partial^2 \rho_{\rm B}}{\partial \theta^2}+i(2\overline{|\rho_F|^2}+|\rho_B|^2)\rho_{\rm B}+i\beta\overline{\rho_{\rm F}},\\
    &\frac{d\rho_{\rm L}}{d\tau}=i(\Delta_{\rm L}-\Delta)\rho_{\rm L}-\Gamma \rho_{\rm L}+G_{\rm L}(1+i\alpha_g)\rho_{\rm L}-2\sqrt{T\eta\Lambda}e^{i\phi_{\rm B}}\overline{\rho_{\rm B}},
    \label{Eq:eqom}
    \end{split}
\end{equation}
where normalized coefficients $\tau=\kappa t/2$, $\Delta=2\delta\omega/\kappa$, $d_2=D_2/\kappa$, $\Delta_{\rm L}=2\delta\omega_{\rm L}/\kappa$, $\eta=\kappa_{\rm R}/\kappa$, $\Lambda=\kappa_{\rm L}/\kappa$, $\Gamma=\gamma_{\rm L}/\kappa$ and $G_{\rm L}=g_{\rm L}/\kappa$. The normalized pump term $F=-2\sqrt{T\eta\Lambda}e^{i\phi_{\rm B}}\rho_{\rm L}$. Note that if the $Q$ of the laser diode is limited by the coating on its emission end, then $\Lambda=Q_{\rm m}/Q_{\rm d}$ with $Q_{\rm m}$ the $Q$ of the microresonator and $Q_{\rm d}$ the $Q$ of the laser diode.

Expanding the laser field as $\rho_{\rm L}=|\rho_{\rm L}|e^{i\phi_{\rm L}}$, the dynamics of the amplitude and phase are given by
\begin{equation}
    \frac{1}{|\rho_{\rm L}|}\frac{d|\rho_{\rm L}|}{d\tau}=G_{\rm L}-\Gamma-{\rm Re}[2\sqrt{T\eta\Lambda}e^{i\phi_{\rm B}}\frac{\overline{\rho_{\rm B}}}{\rho_{\rm L}}],
\end{equation}
\begin{equation}
    \frac{d\phi_{\rm L}}{d\tau}=\Delta_{\rm L}-\Delta+\alpha_g G_{\rm L}-{\rm Im}[2\sqrt{T\eta\Lambda}e^{i\phi_{\rm B}}\frac{\overline{\rho_{\rm B}}}{\rho_{\rm L}}].
    \label{Eq:phiL}
\end{equation}
In the presence of sufficient gain saturation, the laser dynamics can be adiabatically eliminated so that $d|\rho_{\rm L}|/d\tau=0$. Therefore the gain can be solved as
\begin{equation}
    G_{\rm L}=\Gamma+{\rm Re}[2\sqrt{T\eta\Lambda}e^{i\phi_{\rm B}}\frac{\overline{\rho_{\rm B}}}{\rho_{\rm L}}].
    \label{Eq:gain}
\end{equation}
Substituting Eq. \ref{Eq:gain} into Eq. \ref{Eq:phiL}, we obtain
\begin{equation}
\begin{split}
    \frac{d\phi_{\rm L}}{d\tau}&=\Delta_{\rm L}-\Delta+\alpha_g\Gamma-{\rm Im}[2(1-i\alpha_g)\sqrt{T\eta\Lambda}e^{i\phi_{\rm B}}\frac{\overline{\rho_{\rm B}}}{\rho_{\rm L}}]\\&=\Delta_{\rm L}-\Delta+\alpha_g\Gamma+4T\eta\Lambda\sqrt{1+\alpha_g^2}\:{\rm Im}[e^{i\psi}\frac{\overline{\rho_{\rm B}}}{F}],
    \label{Eq:adler}
    \end{split}
\end{equation}
where 
\begin{equation}
    \psi=2\phi_{\rm B}-{\rm arctan}(\alpha_g).
\end{equation}
Note that Eq. \ref{Eq:adler} resembles the form of Adler's equation \cite{adler1946study}, whose stationary solution gives the frequency of the self-injection locked laser as
\begin{equation}
    \Delta=\Delta_{\rm L}+\alpha_g\Gamma+K\:{\rm Im} [e^{i\psi}\frac{\overline{\rho_{\rm B}}}{F}].
    \label{Eq:detuning}
\end{equation}
It should be noted that, the locking strength $K=4T\eta\Lambda\sqrt{1+\alpha_g^2}$, is usually much greater than 1 ($>10^3$ in this work).

The linewidth-reduction factor is derived as follows. For simplicity, we ignore all nonlinear terms by assuming homogeneous intracavity field, i.e., $\overline{\rho_{\rm F,B}}=\rho_{\rm F,B}$. Therefore, at steady state, the backscattered field is given by
\begin{equation}
    \rho_{\rm B}=\frac{i\beta F}{(1+i\Delta)^2+|\beta|^2}.
    \label{Eq:ABAL}
\end{equation}
Substituting Eq. \ref{Eq:ABAL} into Eq. \ref{Eq:detuning}, we have
\begin{equation}
    \Delta=\Delta_{\rm L}+\alpha_g\Gamma+K\:{\rm Im} [ie^{i\psi}\frac{\beta }{(1+i\Delta)^2+|\beta|^2}]=\Delta_{\rm L}+\alpha_g\Gamma+K\chi.
\end{equation}
The equation has two solutions. However, stable self-injection locking is established only when $\partial\chi/\partial\Delta<0$. Assuming the system is operating at steady state, the linewidth-reduction factor at low offset frequencies can be calculated using
\begin{equation}
    \alpha=(\frac{\partial\Delta_{\rm L}}{\partial\Delta})^2,
\end{equation}
which can be solved analytically in the two following regimes:

\noindent{\bf Weak backscattering ($\beta\ll 1$):} In this case, the mode will remain as a singlet. Stable locking occurs when $\psi\approx-\pi/2$. Assuming the laser is locked to the center of the mode, i.e., $\Delta=0$, the linewidth-reduction factor yields
\begin{equation}
    \alpha\approx 64(1+\alpha_g^2)T^2\eta^2\Lambda^2|\beta|^2=64(1+\alpha_g^2)T^2\eta^2|\beta|^2\frac{Q_{\rm R}^2}{Q_{\rm d}^2},
    \label{Eq:noiseweak}
\end{equation}
where $Q_{\rm R}$ and $Q_{\rm d}$ stand for the $Q$ of the microresonator and the laser diode, respectively.

\noindent{\bf Strong backscattering ($\beta\gg 1$):} In this case, the mode will split as doublets. When $\psi\approx 0 (\pi)$, the laser will be locked to the mode at the red (blue) side. Assuming the laser is locked to the center of a split mode, i.e., $\Delta=\pm |\beta|$, the linewidth-reduction factor yields
\begin{equation}
    \alpha\approx 4(1+\alpha_g^2)T^2\eta^2\Lambda^2=4(1+\alpha_g^2)T^2\eta^2\frac{Q_{\rm R}^2}{Q_{\rm d}^2},
\end{equation}
which is irrelevant to the backscattering coefficient $\beta$. 

Typical values of these parameters in our systems are: $\alpha_g=2.5$, $T=-6$ dB, $\eta=0.5$, $Q_{\rm d}=10^4$. For mode featuring loaded $Q$ of 50~M and split resonances, the maximum estimated noise reduction factor is around 70~dB, which is 20 dB higher than the noise suppression achieved in experiment. Under single-mode operation, we expect the locking point to be detuned from resonance center to avoid nonlinearity.

\section{Theory: spectral dependence of noise suppression factor}

In this section, we derive the spectral dependence of the noise suppression factor, where the frequency response of the microresonator is taken into consideration. For simplicity, the backscattering is assumed weak enough to not cause mode splitting ($\beta\ll1$). The laser is assumed locked to the center of the resonance, i.e., $\Delta=0$. Nonlinearities, dispersion and the phase-amplitude coupling coefficient $\alpha_g$ are also ignored. The steady state solutions from Eq. \ref{Eq:eqom} now read
\begin{equation}
    \begin{split}
        &\rho_{\rm F}=R\rho_{\rm L},\\
        &\rho_{\rm B}=i\beta R\rho_{\rm L},\\
    \end{split}
    \label{Eq:steadyfield}
\end{equation}
where $R=-2\sqrt{T\eta\Lambda}e^{i\phi_{\rm B}}$. The fluctuation of laser can be introduced using a Langevin term $f$, which includes spontaneous emission and carrier density fluctuations. As a result, the system will fluctuate in the vicinity of the steady state, as denoted by field perturbation $u_{\rm F}$, $u_{\rm B}$ and $u_{\rm L}$. With proper linearization, the dynamics of perturbation terms are formulated as
\begin{equation}
\begin{split}
    &\Dot{u_{\rm F}}=-u_{\rm F}+i\beta u_{\rm B}+Ru_{\rm L},\\
    &\Dot{u_{\rm B}}=-u_{\rm B}+i\beta u_{\rm F},\\
    &\Dot{u_{\rm L}}=i\Delta_{\rm L}u_{\rm L}+(G_{\rm L}-\Gamma)u_{\rm L}-2\epsilon G_{\rm L} A_u \rho_{\rm L}+R u_{\rm B}+f(t),
    \label{Eq:eqom_perturbation}
    \end{split}
\end{equation}
where
\begin{equation}
    u_{\rm L}=\rho_{\rm L}(1+A_u)e^{i\phi_u}-\rho_{\rm L}\approx\rho_{\rm L}A_u+i\rho_{\rm L}\phi_u.
\end{equation}
The gain saturation effect is described by $G_{\rm L}=G_o/(1+\epsilon|\rho_{L}|^2)$. The amplitude and phase perturbations of the laser can be studied separately as
\begin{equation}
    \begin{split}
        &\Dot{A_u}=-2\Delta_{\rm L}\phi_u+(G_{\rm L}-\Gamma)A_u-2\epsilon G_{\rm L} A_u +\frac{1}{2}(\frac{R u_{\rm B}+f}{\rho_{\rm L}}+{\rm c.c.}),\\
        &\Dot{\phi_u}=\Delta_{\rm L}A_u+(G_{\rm L}-\Gamma)\phi_u+\frac{1}{2i}(\frac{R u_{\rm B}+f}{\rho_{\rm L}}-{\rm c.c.}).\\
    \end{split}
\end{equation}
Assuming the gain saturation is large, i.e., $\epsilon\gg1$, the amplitude fluctuation diminishes ($A_u\approx 0$). Therefore, using Eq. \ref{Eq:gain} and Eq. \ref{Eq:steadyfield} we obtain
\begin{equation}
    \Dot{\phi_u}=-\frac{i\beta (R^2-R^{*2})}{2}\phi_u+\frac{1}{2i}(\frac{R u_{\rm B}+f}{\rho_{\rm L}}-{\rm c.c.}).
\end{equation}
Applying Fourier transform, the spectral density of $\phi_u$ yields
\begin{equation}
    i\omega\widetilde{\phi_u}=-\frac{i\beta (R^2-R^{*2})}{2}\widetilde{\phi_u}+\frac{1}{2i}(R\widetilde{\frac{ u_{\rm B}}{\rho_{\rm L}}}-R^*\widetilde{\frac{ u^*_{\rm B}}{\rho^*_{\rm L}}})+\frac{1}{2i}(\widetilde{\frac{ f}{\rho_{\rm L}}}-\widetilde{\frac{ f^*}{\rho^*_{\rm L}}}).
    \label{Eq:phiu}
\end{equation}
The spectrum of $u_{\rm B}$ can be derived from Eq. \ref{Eq:eqom_perturbation} as
\begin{equation}
    \begin{split}
    &i\omega\widetilde{u_{\rm F}}=-\widetilde{u_{\rm F}}+i\beta \widetilde{u_{\rm B}}+iR\rho_{\rm L}\widetilde{\phi_u},\\
    &i\omega\widetilde{u_{\rm B}}=-\widetilde{u_{\rm B}}+i\beta \widetilde{u_{\rm F}},\\
    \end{split}
\end{equation}
where $\widetilde{u_{\rm L}}\approx i\rho_{\rm L}\widetilde{\phi_u}$ is used.
\begin{figure}
\centering
\includegraphics[width=0.6\linewidth]{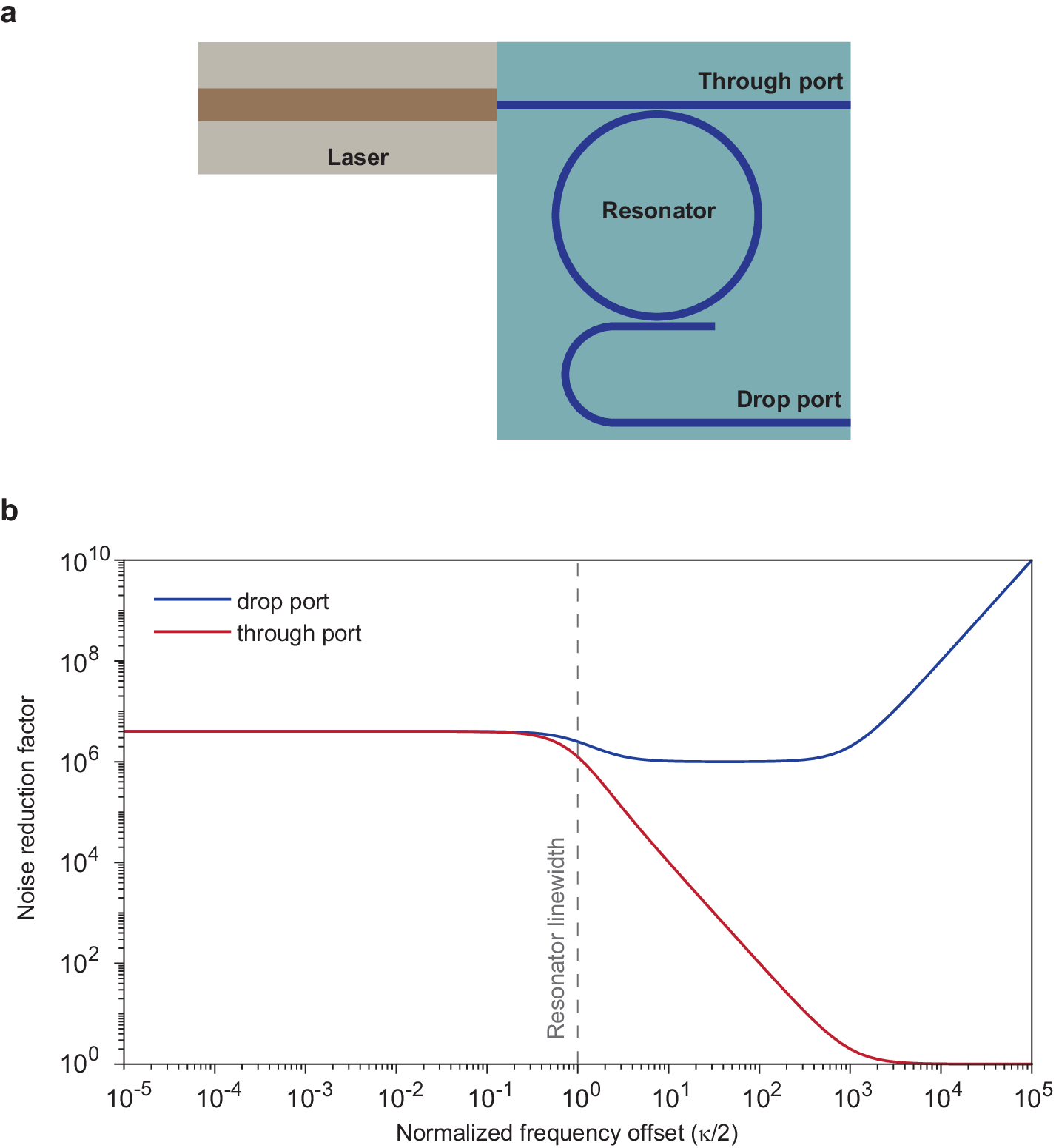}
\caption{{\bf Spectral dependence of noise reduction factor.} {\bf a,} Configuration of a laser butt-coupled to a microresonator with add-drop port. {\bf b,} Calculated noise reduction factor of laser emission from through (red) and drop (blue) port. The frequency offset is normalized to $\kappa/2$, one half of resonator linewidth. The term $4T\eta\Lambda$ is set to be 1000 in the plot.}
\label{figureS4}
\end{figure}
Therefore, we have
\begin{equation}
    \widetilde{u_{\rm B}}=-\frac{1}{(1+i\omega)^2}\beta R\rho_{\rm L}\widetilde{\phi_u},
    \label{Eq:ub}
\end{equation}
where $\beta\ll1$ is applied. Similarly, by taking conjugation and replacing $\omega$ with $-\omega$, the spectrum of $u^*_{\rm B}$ is given by
\begin{equation}
    \widetilde{u^*_{\rm B}}=-\frac{1}{(1+i\omega)^2}\beta R^*\rho^*_{\rm L}\widetilde{\phi_u}.
    \label{Eq:ubdag}
\end{equation}
Substituting Eqs. \ref{Eq:ub} and \ref{Eq:ubdag} into Eq. \ref{Eq:phiu}, we have
\begin{equation}
    i\omega\widetilde{\phi_u}=-\frac{i\beta(R^2-R^{*2})(2i\omega-\omega^2)}{2(1+i\omega)^2}\widetilde{\phi_u}+\widetilde{s},
\end{equation}
where $s=\frac{1}{2i}(\widetilde{\frac{ f}{\rho_{\rm L}}}-\widetilde{\frac{ f^*}{\rho^*_{\rm L}}})$. If the feedback phase $\psi=2\phi_{\rm B}=\pi/2$, we have
\begin{equation}
    R^2-R^{*2}=-8iT\eta\Lambda.
\end{equation}
Therefore, the phase noise of the self-injection locked laser can be written as
\begin{equation}
    |\widetilde{\phi_u}|^2=\frac{|\widetilde{s}|^2}{\omega^2|1+\frac{4\beta T\eta \Lambda(2+i\omega)}{(1+i\omega)^2}|^2}.
\end{equation}
Compared with the free-running laser noise (obtained by setting $\beta=0$), the spectrum of noise reduction factor is given by
\begin{equation}
    \alpha(\omega)=|1+\frac{4\beta T\eta \Lambda(2+i\omega)}{(1+i\omega)^2}|^2.
    \label{Eq:spectral}
\end{equation}

It is noted that at low offset frequencies, Eq. \ref{Eq:spectral} resembles the form of Eq. \ref{Eq:noiseweak} except for correction from amplitude-phase coupling term, which is set to be 0 in the derivation. Plotted in Fig. \ref{figureS4} is a typical spectral dependence of the noise reduction factor, which decreases at the rate of $1/\omega^2$ at frequencies exceeding the resonator linewidth. Such ineffectiveness at high-offset frequencies can be resolved by introducing a drop-port to the microresonator. For a laser emitting from the drop-port of the microresonator, the phase is given by
\begin{equation}
    \phi_F={\rm Im}[\frac{u_{\rm F}}{\rho_{\rm F}}]=\frac{1}{2i}(\frac{u_{\rm F}}{\rho_{\rm F}}-\frac{u^*_{\rm F}}{\rho^*_{\rm F}}),
\end{equation}
whose spectrum yields
\begin{equation}
    \widetilde{\phi_{\rm F}}=\frac{1}{2i}(\frac{u_{\rm F}}{\rho_{\rm F}}-\frac{u^*_{\rm F}}{\rho^*_{\rm F}})=\frac{1}{1+i\omega}\widetilde{\phi_u}.
\end{equation}
Therefore, the noise reduction factor of the drop-port emission versus the free-running laser takes the form
\begin{equation}
    \alpha_{\rm drop}(\omega)=(1+\omega^2)|1+\frac{4\beta T\eta \Lambda(2+i\omega)}{(1+i\omega)^2}|^2.
\end{equation}
The additional noise suppression term at high-offset frequencies is attributed to the filtering effect of the high-Q mode. As a result, the noise reduction factor remains nearly constant across a wider frequency span. Stronger noise suppression at higher-frequencies is expected, though a model involving multiple longitudinal modes should be established at offset frequency on the order of the microresonator $FSR$.

\begin{figure}
\centering
\includegraphics[width=\linewidth]{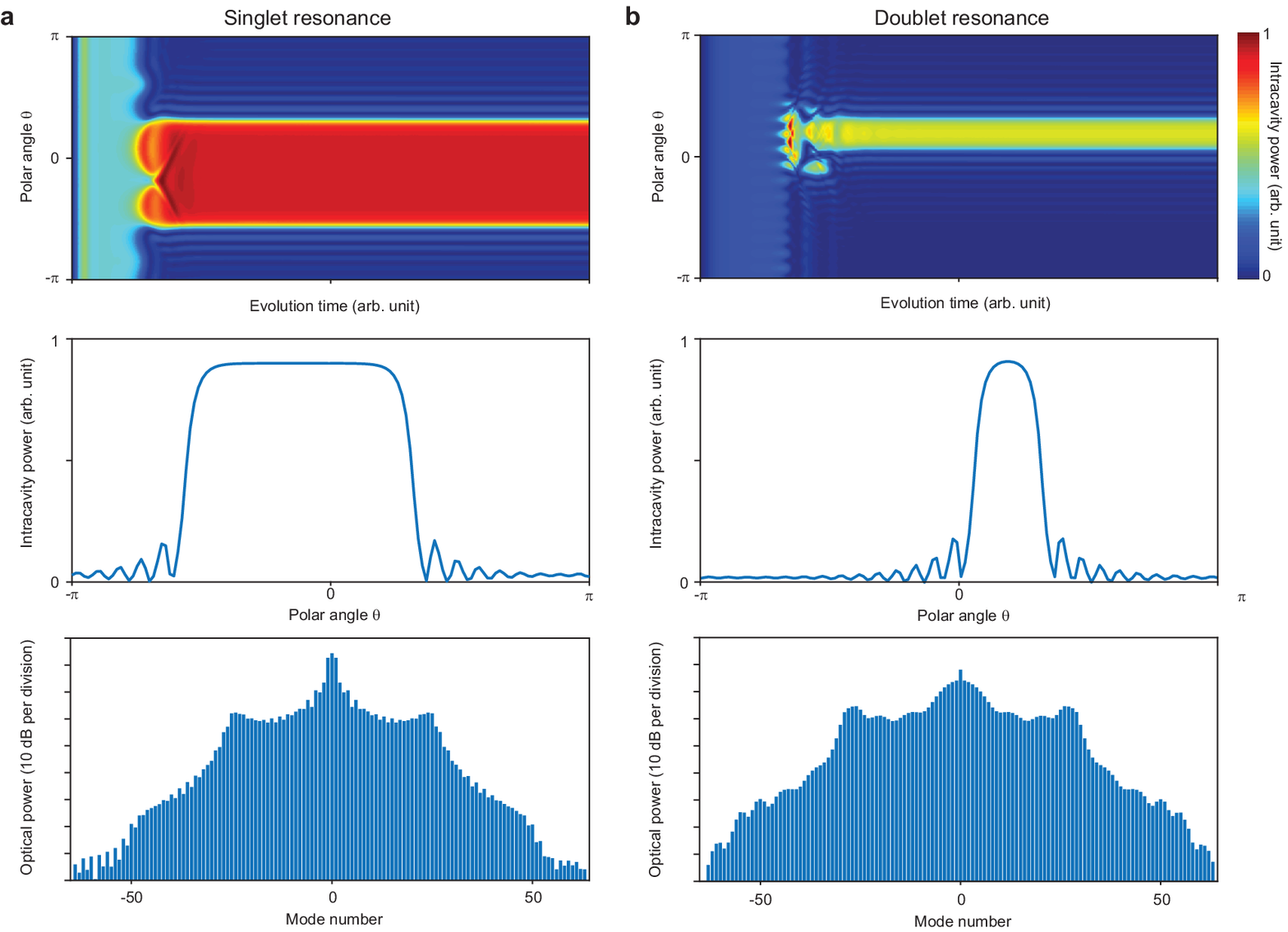}
\caption{{\bf Numerical simulation of Kerr comb formation in microresonators pumped by self-injection locked lasers.} {\bf a,} Evolution of intracavity field (upper panel), steady-state intracavity field (middle panel), and steady-state optical spectrum (lower panel) when the mode is a singlet. The result is identical to that has been provided in Fig. 3 in the maintext. Parameters used in the simulation are: $d_2=-0.01$, $\beta=0.1$, $K=2700$, $F^2=10$, $\psi=-\pi/2$, $\Delta_L+\alpha_g\Gamma=5$. {\bf b,} Evolution of intracavity field (upper panel), steady-state intracavity field (middle panel), and steady-state optical spectrum (lower panel) when the mode splits into doublets. Parameters used in the simulation are: $d_2=-0.01$, $\beta=5$, $K=2700$, $F^2=10$, $\psi=0$, $\Delta_L+\alpha_g\Gamma=5$.}
\label{figureS2}
\end{figure}

\section{Theory: Kerr comb formation}
In this section we use the formalism in Section I to study the dynamics of Kerr comb formation in microresonators pumped by self-injection locked laser. On account of Kerr nonlinearity and dispersion, the intracavity field is no longer homogeneous. Since the $Q$ of the laser cavity is much lower than the microresonator, the dynamics of the laser are much faster than the optical field in the microresonator, which can be assumed to operate at steady state as given by Eq. \ref{Eq:detuning}. Therefore we can retrieve a set of coupled Lugiato-Lefever equations \cite{lugiato1987spatial,shen2020integrated} as
\begin{equation}
\begin{split}
    &\frac{\partial\rho_{\rm F}}{\partial\tau}=-(1+i\Delta)\rho_{\rm F}+id_2\frac{\partial^2 \rho_{\rm F}}{\partial \theta^2}+i(2\overline{|\rho_B|^2}+|\rho_F|^2)\rho_{\rm F}+i\beta \overline{\rho_{\rm B}}+F,\\
    &\frac{\partial \rho_{\rm B}}{\partial\tau}=-(1+i\Delta)\rho_{\rm B}+id_2\frac{\partial^2 \rho_{\rm B}}{\partial \theta^2}+i(2\overline{|\rho_F|^2}+|\rho_B|^2)\rho_{\rm B}+i\beta\overline{\rho_{\rm F}},\\
    &\Delta=\Delta_{\rm L}+\alpha_g\Gamma+K\:{\rm Im} [e^{i\psi}\frac{\overline{\rho_{\rm B}}}{F}].
    \label{Eq:LLE}
    \end{split}
\end{equation}
Numerical simulation based on the equations above is shown in Fig. \ref{figureS2}, where two conditions are considered. With weak backscattering, the mode remains a singlet. By setting the feedback phase as $-\pi/2$, a spontaneous pattern forms, leading to a flat-top pulse in the time domain as shown in Fig. \ref{figureS2}a. Such a pulse is usually referred to as a dark soliton pulse \cite{xue2015mode} or platicon \cite{lobanov2015frequency}. The optical spectrum is similar to the 10 GHz Kerr comb shown in Fig. 3C in the maintext. When the backscattering is strong enough to cause mode splitting, spontaneous Kerr comb formation is also feasible by setting feedback phase $\psi\approx 0$, as shown in Fig. \ref{figureS2}b. Compared with the case of weak-scattering, a pulse with shorter duration is formed, which is attributed to the increased effective detuning resulting from mode splitting. The simulated optical spectrum mimics the shape of the 5 GHz Kerr comb shown in Fig. 3c in the maintext. It should be noted that although stable injection can be established with feedback phase of $\pi$, spontaneous Kerr comb formation is forbidden as imposed by the requirement of parametric oscillation.

\section{Additional measurement}
\begin{figure}[!ht]
\centering
\includegraphics[width=\linewidth]{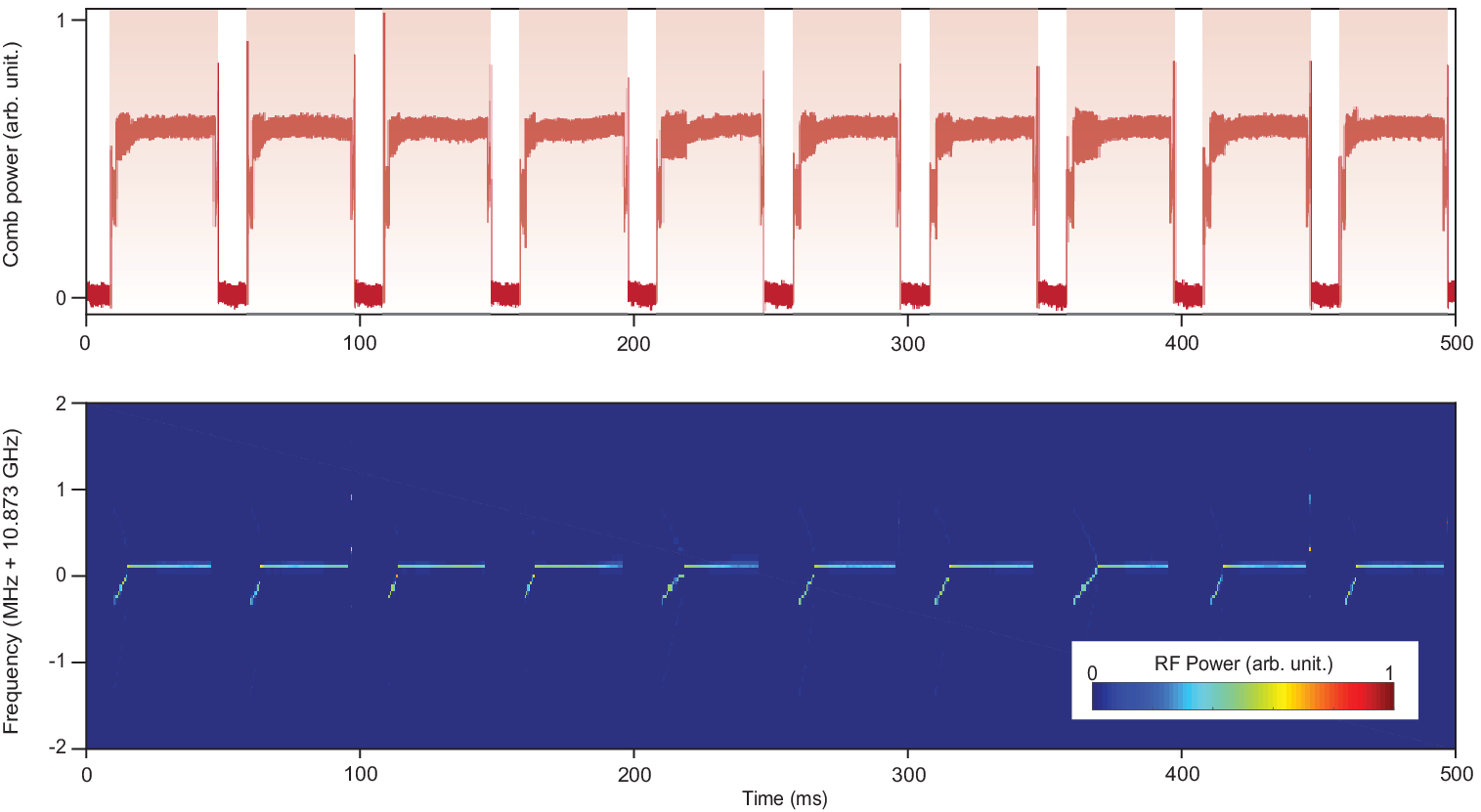}
\caption{{\bf Repeatability of turnkey Kerr comb generation.} Measured comb power (upper panel) and spectrograph of comb repetition rate (lower panel) of 10 consecutive laser switching-on tests. The shaded region indicates the periodic switching-on of lasers.}
\label{figureS3}
\end{figure}

\bibliography{ref2}